\documentclass[useAMS,usenatbib]{mn2e}
\usepackage[dvipdfmx]{graphicx,color}
\usepackage{amsmath}

\title[Bulk Comptonization by Turbulence in Accretion Disks]{Bulk
Comptonization by Turbulence in Accretion Disks}
\author[J. Kaufman and O. M. Blaes]{J. Kaufman\thanks{E-mail:
jason.kaufman09@gmail.com (JK); blaes@physics.ucsb.edu (OMB)}
and O. M. Blaes\\
Department of Physics, University of California, Santa Barbara, CA 93106, USA}
\begin{document}

\date{Accepted ---. Received ---; in original form ---}

\pagerange{\pageref{firstpage}--\pageref{lastpage}} \pubyear{2015}

\maketitle

\label{firstpage}

\begin{abstract}
Radiation pressure dominated accretion discs around compact objects may
have turbulent velocities that greatly exceed the electron thermal velocities within the disc.  Bulk Comptonization by the turbulence may therefore dominate over thermal Comptonization in determining the emergent spectrum. Bulk Comptonization by divergenceless turbulence is due to  radiation viscous dissipation only. It can be treated as thermal Comptonization by solving the Kompaneets equation with an equivalent ``wave" temperature, which is a weighted sum over the power present at each scale in the turbulent cascade. Bulk Comptonization by turbulence with non-zero divergence is due to both pressure work and radiation viscous dissipation. Pressure work has negligible effect on photon spectra in the limit of optically thin turbulence, and in this limit radiation viscous dissipation alone can be treated as thermal Comptonization with a temperature equivalent to the full turbulent power. In the limit of extremely optically thick turbulence, radiation viscous dissipation is suppressed, and the evolution of local photon spectra can be understood in terms of compression and expansion of the strongly coupled photon and gas fluids. We discuss the consequences of these effects for self-consistently resolving and interpreting turbulent Comptonization in spectral calculations in radiation MHD simulations of high luminosity accretion flows.
\end{abstract}


\section{Introduction}
\label{sec_intro}

Electron scattering is one of the most important processes in determining the
emergent thermal spectrum from models of optically thick accretion discs
around black holes and neutron stars.  Electron scattering opacity generally
dominates absorption opacity in the atmospheres of the innermost regions of such
discs \citep{sha73}.  In the case where coherent (Thomson) scattering is a
good approximation, the resulting local thermal spectrum of some annulus
in the disc is generally harder than a blackbody with the same effective
temperature, due to incomplete thermalization at the scattering photosphere.
However,
%provided the black hole mass is not too large or the accretion rate
%is not too high,
incoherent (Compton) scattering in the disc surface layers by thermal electrons
can reduce this spectral hardening by increasing the energy exchange between
the photons and the plasma \citep{shi95,dav05,dav06}.

In addition to the disc atmosphere itself, many models invoke a powerful
corona above the disc consisting of high temperature or nonthermal electrons that Compton upscatter disc photons
to produce the energetically significant hard X-rays that exist in certain
classes of active galactic nuclei and in certain black hole X-ray binary
accretion states (e.g. \citealt{haa91,sve94}).
These hard X-rays in turn interact with the relatively cool disc atmosphere
to produce reflection spectra that are widely observed in many black
hole sources (e.g. \citealt{lig88,ros93}).

In this work we explore turbulent Comptonization, which is the effect of Comptonization by bulk, turbulent electron motions on photon spectra \citep{soc04,soc10}. In sources with radiation pressure dominated accretion flows, bulk velocities may exceed thermal electron velocities, a phenomenon first pointed out in \cite{soc04}. The turbulent speeds $v_{\rm turb}$ on the outer scale of an MHD turbulent cascade will be of order the Alfv\'en speed, and the ratio of this
to the root mean square electron thermal velocity is therefore
\begin{equation}
{v_{\rm turb}\over\langle v_{\rm e}^2 \rangle^{1/2}}\sim
\left({P_{\rm mag}\over P_{\rm rad}}\right)^{1/2}
\left({P_{\rm rad}\over P_{\rm gas}}\right)^{1/2}
\left(m\over m_{\rm p}\right)^{1/2}.
\end{equation}
Here $P_{\rm gas}$, $P_{\rm rad}$, and $P_{\rm mag}$ are the gas, radiation,
and magnetic pressures, respectively, and $m_{\rm p}/m$ is the
ratio of the proton to electron mass.
%More precisely, there should be a sqrt(2/3/mu) factor here, but this is
%very close to unity as mu=0.6.
Stratified shearing box simulations of magnetorotational turbulence
generally have disc atmospheres that are supported by magnetic fields rather
than thermal pressure \citep{mil00,hir06,hir09,gua11,jia14a}.
Hence the first factor generally exceeds unity in an
otherwise radiation pressure dominated disc.  Bulk speeds on the outer
scale of the turbulence will therefore exceed the electron thermal speeds
whenever the radiation pressure to gas pressure ratio exceeds the ratio
of the proton to electron mass ratio, and even smaller depending on how
magnetically supported is the disc atmosphere.  In this regime,
bulk Comptonization by the turbulence may dominate thermal Comptonization
in determining the shape of the spectrum emitted by a local patch
of the disc.

This regime is commonly reached for near Eddington accretion
on black holes of all mass scales.  Indeed, the inner disc solution
of the standard geometrically thin model of \citet{sha73} gives vertically
averaged radiation to gas pressure ratios of approximately
\begin{equation}
{P_{\rm rad}\over P_{\rm gas}}\sim10^7\alpha^{1/4}\left({M\over10^8M_\odot}
\right)^{1/4}\eta^{-2}\left({L\over L_{\rm Edd}}\right)^2
\left({r\over r_{\rm g}}\right)^{-21/8},
\end{equation}
where $\alpha$ is the ratio of turbulent stress to thermal pressure, $M$ is
the black hole mass, $\eta\equiv L/(\dot{M}c^2)$ is the radiative efficiency
of the disc as a whole, $L/L_{\rm Edd}$ is the luminosity in Eddington
units, and $r/r_{\rm g}$ is the radius in the disc in units of the
gravitational
radius $r_{\rm g}\equiv GM/c^2$.  Hence the radiation to gas pressure ratio
of the innermost disc will generally exceed the proton to electron mass
ratio for near-Eddington accretion even for stellar mass black holes, and
certainly for supermassive black holes.  On the other hand, energy
exchange between the photons and the plasma is generally dominated by
true absorption opacity in standard disc models
for the most supermassive black holes \citep{lao89,hub01}, which may reduce bulk Comptonization by turbulence in these sources.

We are also motivated by observational
evidence in many sources for a cooler, more
optically thick medium that also Compton upscatters disc photons in high luminosity sources, but not
to the high energies usually associated with the more traditional external corona.
The soft X-ray excess of the type 1 Seyfert NGC~5548 was fit by \citet{mag98}
with a warm ($k_{\rm B}T\sim0.3$~keV) optically thick ($\tau\sim30$)
Comptonizing medium.
Similar parameters were obtained for the soft X-ray excess of the type 1
Seyfert Mrk 509 by \citet{meh11}: $k_{\rm B}T\simeq0.2$~keV
and scattering optical depth $\tau\simeq16$ to $18$.
\citet{don12} fit the soft X-ray excess of the extreme Narrow Line Seyfert 1
RE~J0134+396 assuming extra Comptonization within the disc atmosphere itself
with $k_{\rm B}T\simeq0.23$~keV and $\tau\simeq11$.  Such a spectral
decomposition is consistent with the lack of short time scale variability
in the soft X-ray excess compared to the hard X-rays in this source
\citep{mid09}.  \cite{col15} fit the soft X-ray excesses of
eleven bright quasars with Comptonization in a medium with
$k_{\rm B}T=0.2$~keV
and $\tau=10$.  \citet{kol11} fit soft intermediate state spectra of
the black hole X-ray binary GX 339-4 with a Comptonizing medium with
$k_{\rm B}T\simeq0.7$~keV and $\tau\simeq30$.  Thermal Comptonization
fits to the soft X-ray spectra of a number of ultraluminous X-ray sources
have $k_{\rm B}T\simeq1-3$~keV
and $\tau\simeq6-80$ \citep{sto06,gla09,yos13}.
Such fits have been nicely confirmed with {\it NuSTAR} and
{\it XMM-Newton} spectra of the ultraluminous X-ray sources
NGC 1313 X-1 and X-2, which require warm $k_{\rm B}T\simeq1-2$~keV,
optically thick $\tau\simeq10-16$ Comptonizing media \citep{bac13}.
\citet{gla09} and \citet{yos13} suggest that one way of forming these
optically thick cool coronae is by a radiation pressure driven outflow.
The alternative we consider here is bulk Comptonization by turbulence in the disc atmosphere itself.

Comptonization by bulk motions in the accretion flow has also been considered by others.  \citet{bla81a,bla81b} considered bulk Comptonization in converging flows and shocks.  Starting from this seminal work, bulk
Comptonization by radial flows has been calculated in detail by numerous
authors \citep{pay81,col88,tit97,psa01,nie06}.
\citet{kaw12} included bulk Comptonization in their Monte Carlo
calculations of photon spectra from radiation MHD simulations of
super-Eddington accretion flows, and found that it produced significant
spectral hardening which resembled spectra of ultra-luminous X-ray sources.
Here we focus on smaller scale bulk Comptonization by turbulence within the disc atmosphere itself.
Turbulent Comptonization has also been invoked in other areas of astrophysics.
\citet{zel72} and \cite{cha75} used then current limits on cosmic microwave background
temperature anisotropies to constrain possible turbulent energy on cosmological
scales prior to recombination.  \citet{tho94} considered Comptonization
by Alfv\'enic turbulence in a relativistic outflow as a model for the
spectrum of gamma-ray bursts.

The purpose of this paper is to determine how photon spectra produced by turbulent Comptonization depend on properties of the turbulence itself, and how to resolve and interpret this effect in radiation MHD simulations. The structure of this paper is as follows. In section \ref{sec_general} we show that the macroscopic physical origins of turbulent Comptonization energy exchange are work due to radiation pressure and viscous dissipation due to the radiation viscous stress tensor, and we discuss why this requires us to treat divergenceless turbulence separately from turbulence with non-zero divergence. In section \ref{sec_energy_exchange} we discuss the consequences of this for correctly implementing radiative transport in simulations, and derive the appropriate radiation energy equation in both lab and fluid frame variables. In section \ref{sec_incompressible} we address the conjecture of \cite{soc04} that turbulent Comptonization can be treated as thermal Comptonization by solving the Kompaneets equation with an equivalent ``wave" temperature critically dependent on the photon mean free path. We show this is true only for divergenceless turbulence, derive the exact wave temperature with an analytic solution of the radiative transfer equation, and use this result to discuss how the wave temperature depends on the power spectrum of the turbulence. To provide physical insight, we also perform an intuitive, heuristic calculation of the wave temperature which well approximates the analytic solution. In section \ref{sec_compressible} we consider bulk Comptonization by turbulence with non-zero divergence. We show that Comptonization by turbulence whose wavelengths are short relative to the photon mean free path can be treated as thermal Comptonization with an equivalent temperature given by the full turbulent power. In the limit of extremely optically thick turbulence, we show how the evolution of local photon spectra can be understood in terms of compression and expansion of the strongly coupled photon and gas fluids. In section \ref{sec_discussion} we discuss our results, and we summarize our findings in section \ref{sec_conclusion}.

\section{General considerations of turbulent Comptonization}
\label{sec_general}

In order to determine how photon spectra produced by turbulent Comptonization depend on properties of the turbulence itself, it is useful to understand the distinct macroscopic physical origins of net energy exchange due to this effect. We show that these are simply the work done by radiation pressure and radiation viscous dissipation, and discuss the major consequences of this. We limit our consideration in this paper to non-relativistic velocities.

Before proceeding, we define terms and quantities that will be used repeatedly. By bulk Comptonization we mean Comptonization by bulk motions in general, and by turbulent Comptonization we mean bulk Comptonization specifically due to turbulence. We denote the characteristic photon mean free path $\lambda_{\rm p} = (n_{\rm e} \sigma_{\rm T})^{-1}$, where $n_{\rm e}$ is the electron density and $\sigma_{\rm T}$ is the Thompson cross section. We denote the typical length scale for bulk velocity variations $\lambda \equiv 2\pi/k$, such as the wavelength if there is a well-defined spatial period. Unless otherwise stated, by the terms
optically thin and thick we mean $\lambda_{\rm p} \gg \lambda_{\max}$ and $\lambda_{\rm p} \ll \lambda_{\min}$, where $\lambda_{\rm min}$ and $\lambda_{\rm max}$ are the minimum and maximum length scales in the turbulent cascade, respectively, not referring to the optical depth that a photon would need to travel to escape
the medium.

Net energy exchange due to bulk Comptonization is simply the net energy exchange between gas mechanical energy and radiation. Inside the photosphere, the mechanical energy loss rate per unit volume to radiation is
\begin{align}
\phi = P^{ij} \partial_i v_j,
\label{eq_phi}
\end{align}
where $P^{ij}$ is the lab frame radiation pressure tensor. This can also be written as
\begin{align}
\label{energy_exchange}
\phi = P {\bf \nabla} \cdot {\bf v} + P_{{\rm vis,shear}}^{ij} D_{ij},
\end{align}
where $P = P^{ii}/3$ is the trace of the radiation pressure tensor, $P_{{\rm vis,shear}}^{ij} = P^{ij} - P\delta^{ij}$ is the radiation viscous shear stress tensor, and 
\begin{align}
D_{ij} = \frac{1}{2}\left(\partial_i v_j + \partial_j v_i \right) - \frac{1}{3} {\bf \nabla} \cdot {\bf v} \delta_{ij}
\label{eq_vel_shear}
\end{align}
is the velocity shear tensor. We see that the energy exchange is separated into two pieces, one due to only a diverging velocity field and another due to a shearing velocity field in the presence of a radiation viscous shear stress tensor. The first piece has contributions from two effects, ordinary work done by radiation pressure, and radiation viscous dissipation. The former effect is first order in velocity since it is due to the contribution to $P$ that is zeroth order in velocity. Energy exchange due to viscous effects, on the other hand, is second order in the velocity, as $P_{{\rm vis,shear}}^{ij}$ and the relevant contribution to $P$ must themselves be at least first order in velocity since they are a consequence of the velocity field.

\cite{soc04} conjectured that turbulent Comptonization can be treated as thermal Comptonization by solving the Kompaneets equation with an equivalent ``wave" temperature critically dependent on the photon mean free path. This is physically intuitive for divergenceless turbulence since in this case energy exchange is entirely due to radiation viscous dissipation and is therefore second order in velocity. In section \ref{sec_incompressible} we prove this conjecture for an arbitrary, divergenceless velocity field of uniform density, derive the exact expression for the wave temperature, and show how it is related to the standard expression for radiation viscosity in the optically thick limit. The connection between bulk Comptonization and the radiation viscous shear stress tensor is useful because it correctly suggests that the wave temperature can be written in terms of $P_{{\rm vis,shear}}^{ij}$. Since pressure work, on the other hand, is an effect first order in velocity, and since Comptonization by a velocity field with non-zero divergence is a combination of pressure work and radiation viscous dissipation, it is not surprising that in this case bulk Comptonization cannot be treated as thermal Comptonization, as we show in section \ref{sec_compressible}.

But in the optically thin limit, i.e. when the mean free path is significantly larger than the largest length scale in the turbulence, energy exchange that is first order in velocity vanishes since photons are equally likely to downscatter as they are to upscatter. Bulk Comptonization by a velocity field with non-zero divergence is then solely due to radiation viscous dissipation, and in section \ref{sec_compressible} we show that it may be treated as thermal Comptonization by solving the Kompaneets equation with a wave temperature equivalent to the full bulk power.

In the optically thick case, i.e. when the mean free path is significantly smaller than the smallest scale in the turbulence, the lowest order energy exchange is the work done by radiation pressure to compress the gas, since it is first order in velocity and since radiation viscous effects are suppressed by a factor $\lambda_{\rm p}^2/\lambda^2$ (section \ref{sec_incompressible}). \cite{soc04} assumed that effects first order in velocity always vanish on average for turbulent eddies, but in the optically thick limit photons trapped in a converging (diverging) region undergo systematic upscattering (downscattering). In the extremely optically thick limit in which the photon and gas fluids are strongly coupled, velocity convergence corresponds to compression in which gas mechanical energy is transferred locally to the photons. In section \ref{sec_compressible} we show that in this process a locally thermal photon distribution remains thermal and only changes temperature, completely analogous to the evolution of the cosmic microwave background radiation under the expansion of the universe. Unlike energy exchange due to viscous dissipation, this process is reversible. The effect of this process on the emergent spectrum of the disc will depend primarily on how effectively photons are able to escape from such regions to the observer.

\section{Resolving energy exchange due to bulk Comptonization in radiation MHD simulations}
\label{sec_energy_exchange}
Self-consistent radiation MHD simulations of turbulent, radiation pressure dominated accretion flows now exist, both in local vertically stratified shearing box geometries \citep{hir09,bla11,jia13} and in global simulations \citep{ohs11,tak13,jia14b,mck14,sad13}. Although these simulations use frequency integrated equations, the emergent radiation spectrum can be computed, including the effects of bulk Comptonization, using post-processing Monte Carlo simulations. Indeed, this has already been done by \citet{kaw12}. However, in order for such calculations to be self-consistent, the frequency integrated radiation MHD equations used in the simulations themselves must include energy exchange due to bulk Comptonization. We now discuss the consequences of the macroscopic physical origins of such energy exchange detailed in section \ref{sec_general} for ensuring this effect is captured in simulations. We then proceed to derive the appropriate frequency integrated source terms due to Compton scattering for the gas and radiation energy equations in both lab frame and fluid frame variables. Using these results, we discuss the extent to which bulk Comptonization is captured by existing radiation MHD simulation codes.

The decomposition of bulk Comptonization energy exchange into pressure work and radiation viscous dissipation shows that radiation MHD schemes that neglect contributions to the viscous stress tensor that are first order in velocity cannot capture bulk Comptonization energy exchange due to a shearing velocity field or any optically thin velocity field with non-zero divergence. As these effects are second order in velocity, we also note that a necessary, but not sufficient, condition for capturing these effects is inclusion of energy terms second order in velocity. Without such terms, turbulence in this form, instead of exchanging energy with photons, will eventually cascade down to the gridscale (or viscous or resistive scale if the code has explicit viscosity or resistivity), and increase the internal energy of the gas. Gas internal energy may then be exchanged with photons through thermal Comptonization. The omission of  viscous dissipation by radiation therefore does not prevent the eventual transfer of turbulent energy to radiation, but it may have other physical effects that can in turn affect radiation spectra. 

To derive the appropriate frequency integrated source terms due to Comptonization, we start with the zeroth moment of the radiative transfer equation, correct to order $v^2/c^2$ and $\epsilon/mc^2$ \citep{psa97},
\begin{eqnarray}
{1\over n_{\rm e}\sigma_{\rm T}}\left({1\over c}{\partial n\over\partial t}
+\nabla_in^i\right)=\cr
{1\over\epsilon^2}{\partial\over\partial\epsilon}
\Biggl\{\epsilon^3\Biggl[\frac{\epsilon}{mc^2}n+\left(\frac{k_{\rm B}T_{\rm e}}{mc^2}+{1\over3}\frac{v^2}{c^2}\right)\epsilon
{\partial\over\partial\epsilon}n\cr
+{3\over4}\frac{\epsilon}{mc^2}\left(n^2-n^i n^i+n^{ij} n^{ij}-
n^{ijk}n^{ijk}\right)\Biggr]+\frac{v_i}{c} n^i\Biggr\}\cr
+\left({18\over5}+{17\over5}\epsilon{\partial\over\partial\epsilon}
+{11\over20}\epsilon^2{\partial^2\over\partial\epsilon^2}\right)
\left(n^{ij}{v_iv_j\over c^2}-{v^2\over3c^2}n\right).
\label{eqpsaltis}
\end{eqnarray}
Here $\epsilon$ is the photon energy, and the various angle-averaged moments are defined in terms of the
energy and direction ($\hat{\bf\ell}$) dependent photon occupation number
$n(\epsilon,\hat{\bf\ell})$ by
\begin{eqnarray}
n(\epsilon)\equiv\oint d\Omega n(\epsilon,\hat{\bf\ell}),\cr
n^i(\epsilon)\equiv\oint d\Omega \ell^i n(\epsilon,\hat{\bf\ell}),\cr
n^{ij}(\epsilon)\equiv\oint d\Omega \ell^i\ell^j n(\epsilon,\hat{\bf\ell}),\cr
\,\,\,\,{\rm and}\,\,\,\,
n^{ijk}(\epsilon)\equiv\oint d\Omega \ell^i\ell^j\ell^k
n(\epsilon,\hat{\bf\ell}).
\end{eqnarray}
In principle, the energy equation is obtained by writing Eq. (\ref{eqpsaltis}) in terms of moments of the specific intensity and then integrating over all frequencies. Unfortunately, we cannot integrate over terms multiplied by $\epsilon / m c^2$ without prior knowledge of the spectrum. For the purpose of simulations, then, we make two approximations. First, we observe that the fractional energy change per scattering off of non-relativistic electrons is small, so that only regions inside the photosphere contribute to Comptonization. Since the stimulated scattering terms are already order $\epsilon / m c^2$ and in these regions departures from isotropy are order $v/c$, we make the following approximation for these terms:
\begin{align}
n^2 - n^i n^i + n^{ij} n^{ij} - n^{ijk} n^{ijk} \approx \frac{4}{3}n^2.
\end{align}
Second, we assume that the spectrum can be approximated by a Bose-Einstein distribution with temperature $T_r$. With these approximations, Eq. (\ref{eqpsaltis}) yields
\begin{align}
\partial_t E + \partial_i F^i &= n_e \sigma_T c \left(-\left(\frac{v_i}{c}\right)\frac{F^i}{c}+\left(\frac{v}{c}\right)^2 E \right. \notag \\ 
&+ \left.\left(\frac{v_i}{c}\right)\left(\frac{v_j}{c}\right)P^{ij}+4k_{\rm B}\left(\frac{T_e - T_r}{mc^2}\right)E\right).
\label{integrated1}
\end{align}
This is the correct energy equation, but in order for it to capture second order energy exchange, the substituted value of $F^i$ must be calculated with a moment closure scheme that does not neglect contributions to the radiation viscous stress tensor that are first order in velocity. For example, it is not adequate to calculate $F^i$ by substituting $P^{ij} = (E/3)\delta^{ij}$ into the first moment equation.\footnote{However, since the pressure term in the energy equation is already second order in velocity, for the purposes of capturing bulk Comptonization energy exchange it is acceptable to make the approximation $P^{ij} \approx (E/3)\delta^{ij}$ here.} This is equivalent to flux limited diffusion in the diffusion regime, such as that implemented in \cite{hir09}. These do, however, capture energy exchange due to pressure work in the optically thick regime. To show this, we substitute into Eq. (\ref{integrated1}) the standard closure relation, 
\begin{align}
F^i = -\frac{c}{3 n_e \sigma_T} \partial_i E + \frac{4}{3}v_i E,
\label{integrated_closure}
\end{align}
which gives
\begin{align}
\partial_t E + \partial_i \left( -\frac{c}{3 n_e \sigma_T} \partial_i E + v_i E \right) &= - \frac{1}{3} E \partial_i v_i \notag \\
+ n_e \sigma_T c \left(4k_{\rm B}\left(\frac{T_e - T_r}{mc^2}\right)E\right).
\end{align}
We see that energy exchange second order in velocity is not present. Furthermore, we see that energy exchange due to a converging velocity field is indeed the work done by radiation pressure to compress the gas, $-(1/3) E \partial_i v_i \approx - P \partial_i v_i $.

The M1 closure scheme \citep{lev84}, implemented in, e.g., \cite{sad13}, also captures first order energy exchange but not second order energy exchange. This scheme assumes that there exists a frame in which $P^{ij} = \delta^{ij} E/3$. The lab frame radiation pressure tensor can then be expressed in terms of the energy density and flux \citep{sad13}:
\begin{align}
P^{ij} = \left(\frac{1-\xi}{2}\delta^{ij} + \frac{3\xi - 1}{2}f^i f^j \right) E,
\end{align} 
where $f^i = F^i/E$ and
\begin{align}
\xi = \frac{3+4 f^i f^i}{5+2\sqrt{4 - 3f^i f^i}}.
\end{align}
To show that second order energy exchange is not captured, we consider the case of a non-zero radiation viscous shear stress tensor due to a non-relativistic velocity field in an otherwise homogeneous medium. The lowest order contribution to the flux must be first order in velocity. In this scheme, then, the radiation viscous shear stress tensor is zero to first order in velocity, and hence second order energy exchange, which requires a contribution first order in velocity (Eq. \ref{energy_exchange}), is not captured.

Another way to understand why both flux-limited diffusion and the M1 closure scheme fail to capture second order energy exchange is to observe that they both bridge generic optically thick conditions with optically thin conditions, while optically thin turbulence does not fall into either category. In optically thin turbulence, the turbulence length scales are optically thin ($\lambda_{\rm max} \ll \lambda_{\rm p}$), but conditions are otherwise optically thick ($\lambda_{\rm p}\nabla E/E \ll 1$), since we are far enough inside the photosphere that photons must scatter many times before escaping. It seems that only a more sophisticated approach, such as explicitly solving the transfer equation as done by \cite{jia13}, can capture this effect.

We note that \cite{sad15} add an artificial viscosity to the M1 closure scheme in order to address a numerical problem associated with artificial shocks in their simulations. They assume a kinematic radiation viscosity given by
\begin{align}
\nu_s = 0.1 \left(\frac{E}{\rho c^2}\right)\lambda_{\rm p}c,
\end{align}
which, as they acknowledge, underestimates the actual viscosity in the optically thick limit by a factor of $27/80$ (Eq. \ref{scattering_nu}). Since we show (section \ref{sec_incompressible}) that the viscosity is smallest in the optically thick limit (Eq. \ref{scattering_nu_k}), this artificial viscosity always underestimates the actual radiation shear stress tensor, and therefore only partially captures secord order energy exchange.

For completeness, we also write Eq. (\ref{integrated1}) in terms of fluid frame radiation variables, indicated by subscript zero:
\begin{align}
\partial_t \left( E_0 + 2 \left(\frac{v_i}{c}\right)\left(\frac{F_0^i}{c} \right) \right) + \partial_i \left( F_0^i + v_i E_0 + v_j P_0^{ij} \right) &= \notag \\ 
n_e \sigma_T c \left(-\left(\frac{v_i}{c}\right)\frac{F_0^i}{c}+4k_{\rm B}\left(\frac{T_e - T_r}{mc^2}\right)\right).
\label{integrated2}
\end{align}
Since
\begin{align}
\partial_t \left( 2 \left(\frac{v_i}{c}\right)\left(\frac{F_0^i}{c} \right)  \right) \sim v_i \partial_i \left( 2 \left(\frac{v_i}{c}\right)\left(\frac{F_0^i}{c} \right)  \right) \ll \partial_i F_0^i,
\end{align}
Eq. (\ref{integrated2}) simplifies to
\begin{align}
\partial_t E_0 + \partial_i \left( F_0^i + v_i E_0 + v_j P_0^{ij} \right) &= \notag \\ 
n_e \sigma_T c \left(-\left(\frac{v_i}{c}\right)\frac{F_0^i}{c}+4k_{\rm B}\left(\frac{T_e - T_r}{mc^2}\right)\right).
\label{eq_intergrated2}
\end{align}

\section{Comptonization by divergenceless turbulence}
\label{sec_incompressible}

\cite{soc04} (hereafter S04) conjectured that Comptonization by turbulence can be treated as thermal Comptonization by solving the Kompaneets equation with an equivalent ``wave" temperature given by 
\begin{align}
T_{{\rm w}}(\lambda_{\rm p}) \approx \int_{k = 2\pi / \lambda_{\rm p}}^\infty T_{\rm tot}(k)dk, 
\label{Twave1}
\end{align}
where
\begin{align}
T_{\rm tot}(k) = \frac{1}{\frac{3}{2} k_{\rm B} n_{\rm e}}E(k)
\end{align}
is the temperature corresponding to the total electron energy density $E(k)$ at wavenumber $k$. Equation (\ref{Twave1}) generalizes S04 Eq. (8),
\begin{align}
T_{{\rm w}}(\lambda_{\rm p}) \approx T_{{\rm w}}(\lambda_0) \left(\frac{\lambda_{\rm p}}{\lambda_0}\right)^{2/3}, 
\end{align}
which gives $T_{{\rm w}}(\lambda_{\rm p})$ for a Kolmogorov spectrum, $E(k) \propto k^{-5/3}$, with maximum wavelength $\lambda_0$. Equation (\ref{Twave1}) is a weighting scheme of the form
\begin{align}
T_{{\rm w}}(\lambda_{\rm p}) = \int_{0}^\infty f(k)T_{\rm tot}(k)dk,
\label{eq_scheme_form}
\end{align}
with the weighting function $f(k)$ given by
\begin{align}
f_{S04}(k) = \begin{cases}
1, k \geq 2 \pi / \lambda_{\rm p} \\
0, k < 2 \pi / \lambda_{\rm p}.
\end{cases}
\label{Twave3}
\end{align}
This scheme roughly gives full weight to wavelengths less than $\lambda_{\rm p}$ and zero weight to wavelengths greater than $\lambda_{\rm p}$. For a periodic velocity field, since the modes are discrete, Eq. (\ref{eq_scheme_form}) is more clearly written
\begin{align}
T_{{\rm w}}(\lambda_{\rm p}) = \sum_i f(k_i) T_{\rm tot,i},
\label{Twave2}
\end{align}
where $T_{\rm tot,i}$ is the temperature of mode $i$. For the remainder of this section we use Eq. (\ref{Twave2}) since it is more useful for applications to radiation MHD simulations, but note that all results also hold for Eq. (\ref{eq_scheme_form}), as this is just the continuum limit. We also define $\tau_k = 1/\lambda_{\rm p} k$ = $\lambda/2\pi \lambda_{\rm p}$, the optical depth divided by $2\pi$ across a mode with wavenumber $k$. Eq. (\ref{Twave3}), for example, can then be written,
\begin{align}
f_{S04}(k) = \begin{cases}
1, \tau_k \leq 1 / 2\pi \\
0, \tau_k > 1 / 2\pi .
\end{cases}
\label{Twave4}
\end{align}

\begin{figure}
\includegraphics[width = 84mm]{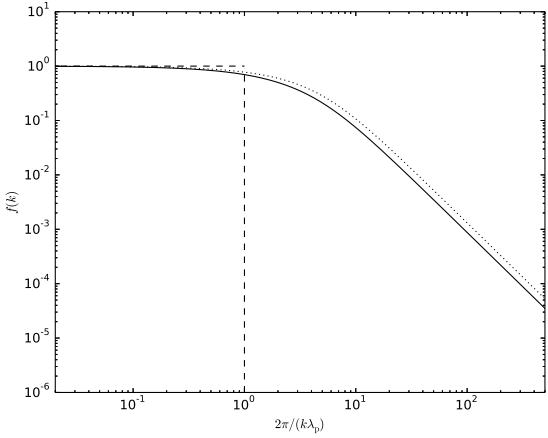}
\caption{Dependence of the mode weighting function on the mode wavelength (in units of the photon mean free path), for a divergenceless velocity field of uniform density. The solid line shows our formal solution, Eq. (\ref{eq:analyticsol}), the dotted line shows our heuristic solution, Eq.
(\ref{eq:heuristicsol}), and the dashed line shows the rough weighting from S04, Eq. (\ref{Twave3}).}
\label{MCdata1}
\end{figure}

The structure of this section is as follows. In section \ref{sec_exact} we build on S04 by proving that Comptonization by an arbitrary velocity field of uniform density can indeed be treated as thermal Comptonization, provided that the field is divergenceless and the photon escape probability is sufficiently small. The exact equivalent wave temperature is of the form Eq. (\ref{Twave2}), with the weighting function $f(k)$ given by 
\begin{align}
f(k) = \frac{2}{\tau_k}\left(\frac{1}{Q(\tau_k)}-
\frac{1}{\tau_k}\right),
\label{eq:analyticsol}
\end{align}
where
\begin{equation}
Q(\tau_k)=\tau_k-{3\over4}\tau_k^3\left[{2\over3}+\tau_k^2-\tau_k(1+\tau_k^2)
\tan^{-1}\left({1\over\tau_k}\right)\right].
\label{eq:qdefinition}
\end{equation}
The limiting cases of $f(k)$ are
\begin{align}
f(k)=\begin{cases}
1 &\mbox{if } \tau_k\rightarrow0\\
{2\over9\tau_k^2} &\mbox{if } \tau_k\rightarrow\infty.\end{cases}
\label{eq_limits}
\end{align}
We plot $f_{S04}(k)$ and $f(k)$ in Fig. \ref{MCdata1}. Since log scaling is used, the curve for $f_{S04}(k)$ disappears for $\lambda > \lambda_{\rm p}$. We see that $f_{S04}(k)$ does roughly approximate $f(k)$. In section \ref{sec_heur}, to provide physical insight into the exact solution, we also find a heuristic, approximate expression for $T_{{\rm w}}(\lambda_{\rm p})$, also plotted in Fig. \ref{MCdata1}, with a simple physical model motivated by ideas put forth in S04. Finally, in section \ref{sec_spec} we use the correct expression for $T_{{\rm w}}(\lambda_{\rm p})$ to discuss how the wave temperature depends on the power spectrum of the turbulence and determine which turbulent wavelengths contribute most to Comptonization.

\subsection{The exact wave temperature}
\label{sec_exact}
To derive Eqs. (\ref{Twave2}) and (\ref{eq:analyticsol}), we first find an expression for the wave temperature in terms of the second moment of the occupation number, since it is the latter which gives rise to the radiation shear stress tensor. We start with the zeroth moment of the radiative transfer equation, Eq. (\ref{eqpsaltis}), correct to order $v^2/c^2$ and $\epsilon/mc^2$. Note that terms of order $\epsilon/mc^2$ are second order in velocity since for Comptonized photons $\epsilon \sim m v^2$. Consider a solution for the angle dependent occupation number $n(\hat{{\bf \ell}},\epsilon,z)$ constructed with the expansion
\begin{align}
n(\hat{{\bf \ell}},\epsilon,z) = \sum_i n_i (\hat{{\bf \ell}},\epsilon,z), 
\label{eq_expansion_1}
\end{align}
where $n_i (\hat{{\bf \ell}},\epsilon,z)$ is formally $i$th order in velocity. We make the ansatz that $n_0 (\hat{{\bf \ell}},\epsilon,z)$ is independent of position and angle, so that to second order in velocity Eq. (\ref{eq_expansion_1}) is
\begin{align}
n(\hat{{\bf \ell}},\epsilon,z) = n_0(\epsilon) + n_1(\hat{{\bf \ell}},\epsilon,z) + n_2(\hat{{\bf \ell}},\epsilon,z).
\label{eq_expansion_2}
\end{align}
To second order in velocity the zeroth moment equation then simplifies to
\begin{align}
\lambda_{\rm p} \left( \frac{1}{c} \partial_t n + \partial_i n ^i\right) = \frac{1}{\epsilon^2} \partial_\epsilon \left( \epsilon^3 \left(\frac{\epsilon}{m c^2} \left( n + n^2 \right) \right. \right. \notag \\
\left. \left. + \left(\frac{k_{\rm B}T_{\rm e}}{mc^2} + \frac{v^2}{3c^2}\right)\epsilon \partial_\epsilon n +\frac{v_i}{c} n^i\right)\right).
\label{eq_zeroth_simplified}
\end{align}
We also invoke the first moment of the transfer equation, \cite{psa97} Eq. (35). We multiply this by $v_i$, so that in steady state, to second order it becomes
\begin{align}
\lambda_{\rm p} \left(\frac{v_i}{c}\right) \partial_j n^{ij} = -\frac{v_i}{c} - \frac{1}{3}\frac{v^2}{c^2} \epsilon \partial_\epsilon n.
\label{eq_first_moment}
\end{align}
Substituting Eq. (\ref{eq_first_moment}) into Eq. (\ref{eq_zeroth_simplified}) gives
\begin{align}
\lambda_{\rm p}\left( \frac{1}{c} \partial_t n + \partial_i n ^i\right) = \frac{1}{\epsilon^2} \partial_\epsilon \left( \epsilon^3 \left(\frac{\epsilon}{m c^2} \left( n + n^2 \right) \right. \right. \notag \\
\left. \left. + \left(\frac{k_{\rm B}T_{\rm e}}{mc^2}\right)\epsilon \partial_\epsilon n - \lambda_{\rm p} \left(\frac{v_j}{c}\right) \partial_i n^{ij}\right)\right).
\end{align}
We average this over space to obtain
\begin{align}
\frac{\lambda_{\rm p}}{c} \partial_t n = \frac{1}{\epsilon^2} \partial_\epsilon \left( \epsilon^3 \left(\frac{\epsilon}{m c^2} \left( n + n^2 \right) \right. \right. \notag \\
\left. \left. + \left(\frac{k_{\rm B}T_{\rm e}}{mc^2}\right)\epsilon \partial_\epsilon n + \lambda_{\rm p}\left\langle \partial_i \left(\frac{v_j}{c}\right) n^{ij}\right\rangle \right)\right).
\end{align}
This is the Kompaneets equation with a bulk Comptonization contribution to the temperature given by
\begin{align}
k_{\rm B}T_{{\rm w}}(\lambda_{\rm p}) =  \frac{\lambda_{\rm p} m c}{2\epsilon \partial_\epsilon n} \left\langle n^{ij}\left(\partial_i v_j + \partial_j v_i \right)\right\rangle.
\label{eq_Twave_n}
\end{align}

To evaluate Eq. (\ref{eq_Twave_n}), we need to derive $n^{ij}$ to first order in velocity. We start with the radiative transfer equation, Eq. (A1) of \cite{psa97}, correct to order $v^2/c^2$ and $\epsilon/mc^2$, and substitute in Eq. (\ref{eq_expansion_2}). In steady state, to first order in velocity, the transfer equation then simplifies to
\begin{align}
\lambda_{\rm p} \ell^i \partial_i n_1(\hat{{\bf \ell}}) = \frac{3}{4}n_1-n_1(\hat{{\bf \ell}})-\ell^i \frac{v_i}{c} \epsilon \partial_{\epsilon} n_0+\frac{3}{4}\ell^i \ell^j  n_1^{ij}.
\label{eq_transfer_first_order}
\end{align}
If the density is constant and the velocity field is divergenceless with sinusoidal mode decomposition
\begin{align}
{\bf v} = \sum_{\bf k} {\bf v_{\bf k}},
\end{align}
then the solution is given by (Appendix \ref{transverse_solution})
\begin{align}
n^{ij} = \frac{1}{3}n \delta^{ij}+ \frac{\lambda_{\rm p} \epsilon \partial_\epsilon n}{3c} \sum_{\bf k} \tau_k^2  f(\tau_k) \left( \partial_i v^j_{\bf k} + \partial_j v^i_{\bf k}\right),
\label{eq_nij}
\end{align}
where $f(k)$ is given by Eq. (\ref{eq:analyticsol}).

Finally, we evaluate Eq. (\ref{eq_Twave_n}) by plugging in Eq. (\ref{eq_nij}) to get
\begin{align}
k_{\rm B}T_{{\rm w}}(\lambda_{\rm p}) &= \notag \\ 
\frac{\lambda_{\rm p}^2 m}{6}&\left\langle \left(\partial_i v^j + \partial_j v^i\right)\sum_{\bf k}\tau_k^2 f(\tau_k)\left(\partial_i v_{\bf k}^j + \partial_j v_{\bf k}^i\right)\right\rangle.
\label{eq_Twave4}
\end{align}
Since the modes are sinusoidal, Eq. (\ref{eq_Twave4}) simplifies to
\begin{align}
k_{\rm B}T_{{\rm w}}(\lambda_{\rm p}) = \sum_{\bf k} \frac{1}{3} m \left\langle v_{\bf k}^2 \right\rangle f(\tau_k),
\end{align}
which is equivalent to Eq. (\ref{Twave2}).

If, however, one solves the Kompaneets equation with an escape probability term, $p_{\rm e} n$, we must also add such a term to the radiative transfer equation, which may affect the occupation number second moment. For an optically thick system, for example, we must require that this term be small compared to the term that sets the diffusion time scale. This term comes from the term $\lambda_{\rm p}  \ell^i \partial_i n({\bf \hat{\ell}})$, which is approximated by
\begin{align}
&\sim \lambda_{\rm p} \ell^i \partial_i \left(n + 3 \ell^j n^j \right) \notag \\
&\sim \lambda_{\rm p} \ell^i \partial_i \left(n - 3 \ell^j\left(\frac{v^j}{3c}\epsilon\partial_{\epsilon}n + \frac{1}{3}\lambda_{\rm p}\partial_j n \right)\right) \notag \\
&\sim \lambda_{\rm p}\left( \ell^i \partial_i n - \ell^i \ell^j \partial_i \left(\frac{v_j}{c} \epsilon\partial_{\epsilon}n\right) - \ell^i \ell^j \lambda_{\rm p}\partial_i \partial_j n \right).
\end{align}
The diffusion timescale is set by the second derivative term,
\begin{align}
\ell^i \ell^j \lambda_{\rm p}^2 \partial_i \partial_j n \sim \left(\frac{\lambda_{\rm p}}{\lambda_{\rm max}}\right)^2 n,
\end{align}
which gives the condition
\begin{align}
p_{\rm e} \ll \left(\frac{\lambda_{\rm p}}{\lambda_{\rm max}}\right)^2.
\label{eq_escape_constraint1}
\end{align}

We now discuss the extent to which, in this case, radiation viscous dissipation can be described by a coefficient of kinematic viscosity $\nu_{\rm s}$, which is usually defined by
\begin{align}
 -P_{\rm vis,shear}^{ij} = \nu_{\rm s} \rho \left( \partial_i v_j+\partial_j v_i \right),
 \label{eq_kinematic}
\end{align}
where $\rho$ is the fluid mass density. First we find the radiation shear stress tensor corresponding to Eq. (\ref{eq_nij}) by frequency integrating and subtracting off the scalar radiation pressure, which gives
\begin{align}
P_{\rm vis,shear}^{ij} = -\frac{4\lambda_{\rm p} E }{3c}\sum_{\bf k} \tau_k^2 f(\tau_k) \left( \partial_i v^j_{\bf k} + \partial_j v^i_{\bf k}\right) .
\label{eq_vis_P}
\end{align}
Incidentally, we note that Eq. (\ref{eq_Twave_n}) can then alternatively be written in terms of the radiation viscous shear stress tensor,
\begin{align}
k_{\rm B}T_{{\rm w}}(\lambda_{\rm p}) =  \frac{-2\lambda_{\rm p} m c}{E} \left\langle P_{\rm vis,shear}^{ij}\left( \partial_i v_j+\partial_j v_i \right)\right\rangle.
\label{eq_Twave_P}
\end{align}
According to Eqs. (\ref{eq_kinematic}) and (\ref{eq_vis_P}), although we cannot in general define a kinematic viscosity coefficient, for any single velocity mode the kinematic radiation viscosity is given by
\begin{align}
\nu_{\rm s, {\bf k}} = \frac{4}{3} \tau_k^2 f(\tau_k) \left(\frac{E}{\rho c^2}\right)\lambda_{\rm p}c,
\label{scattering_nu_k}  
\end{align}
so that the radiation viscous shear stress tensor can be written
\begin{align}
P_{\rm vis,shear}^{ij} = \rho \sum_{\bf k} \nu_{\rm s, {\bf k}} \left( \partial_i v^j_{\bf k} + \partial_j v^i_{\bf k}\right).
\end{align}
However, in the optically thick limit (i.e. $\lambda_{\rm p} \ll \lambda_{\rm min}$), the kinematic viscosity is independent of $\bf k$,
\begin{align}
\nu_{\rm s} = \frac{8}{27} \left(\frac{E}{\rho c^2}\right)\lambda_{\rm p}c,
\label{scattering_nu}
\end{align}
so that in this limit the kinematic viscosity is well-defined for an arbitrary (divergenceless) velocity field. We also note that in this limit Eq. (\ref{eq_Twave4}) agrees with the ``heating temperature" in \cite{cha75}, 
\begin{align}
k_{\rm B} T_{\rm H} = \frac{1}{27}m \lambda_{\rm p}^2 \left(\partial_j v^i + \partial_i v^j\right)^2.
\label{eq_heating_temp}
\end{align}
They also mention that this can be understood in terms of the kinematic radiation viscosity given by Eq. (\ref{scattering_nu}).\footnote{Incidentally, this coefficient for the viscosity was first derived by \cite{mas71}, and it differs by a factor of $\frac{10}{9}$ from the more commonly cited value, $\nu_a = \frac{4}{15}\left(\frac{E}{\rho c^2}\right)c\lambda_{\rm p}$, in, e.g., \cite{wei71}, \cite{wei72}, and \cite{mih84}. The reason for the discrepancy is that the more commonly cited value, first derived by \cite{tho30}, assumes pure absorption, while Eq. (\ref{scattering_nu}) is correct for pure scattering (\citealt{mas71}; \citealt{str76}).}

\subsection{A heuristic wave temperature}
\label{sec_heur} 
To provide physical insight into our analytic solution given by Eqs. (\ref{Twave2}) and (\ref{eq:analyticsol}), we now find a heuristic, approximate expression for $T_{\rm w}(\lambda_{\rm p})$ with a simple physical model motivated by ideas put forth in S04. To do this, we first consider the wave temperature of a single mode with wavenumber $k_i$, which we denote $T_{\rm w}(\lambda_{\rm p}, k_i)$. S04 suggested that photons can only sample turbulent velocities on scales $\lambda \leq \lambda_{\rm p}$, since longer turbulent wavelengths will advect photons back and forth with the flow without allowing the photons to ``sample" their velocities. A rough interpretation of this reasoning leads to 
\begin{align}
T_{\rm w, rough}(\lambda_{\rm p}, k_i) = f_{\rm S04}(k_i)T_{\rm tot,i}.
\end{align}
If one then assumes that the wave temperature of an arbitrary field is the sum of the wave temperatures of its modes, then Eq. (\ref{Twave1}) follows. But a more subtle interpretation of this reasoning suggests that $T_{\rm w}(\lambda_{\rm p}, k_i)$ is determined by the second moment of the distribution of velocity differences between subsequent scatterings, $\langle (\Delta {\bf{v}})^2 \rangle$. In the long wavelength limit, velocity differences between subsequent scatterings are negligible and so the wave power does not contribute to Comptonization. In the short wavelength limit, velocity differences allow photons to sample the full power of the wave. This model of Comptonization suggests that we define
\begin{align}
T_{\rm w, heur}(\lambda_{\rm p}, k_i) = f_{\rm heur}(k_i)T_{\rm tot,i},
\end{align}
where
\begin{align}
f_{\rm heur}(k) \propto \langle (\Delta {\bf{v}})^2 \rangle.
\end{align}

Before proceeding, we note that defining $\langle (\Delta {\bf{v}})^2 \rangle$ is potentially tricky because the distribution of $\Delta {\bf v}$ for a
photon is dependent on its current location, in effect introducing correlations
into subsequent $\Delta {\bf v}$'s. In other words, subsequent $\Delta {\bf v}$'s are not independent.
But if the escape probability is low enough, a condition we quantify below, then the set of $\Delta {\bf v}$'s
encountered by a photon before it escapes is indistinguishable from a set of
$\Delta {\bf v}$'s independently drawn from the position-averaged
$\Delta {\bf v}$ distribution. The order of $\Delta {\bf v}$'s is different in the two cases,
but the total photon energy change does not depend on the order because the
fractional photon energy change per scattering is small for $v^2/c^2 \ll 1$. With these potential problems accounted for, we proceed to calculate $f_{\rm heur}(k)$.

First we find the proportionality constant between $f_{\rm heur}(k)$ and  $\langle (\Delta {\bf{v}})^2 \rangle$ by evaluating both sides in the short wavelength limit. In this limit, the full wave power must contribute, so $f_{\rm heur}(k) \rightarrow 1$. To evaluate $\langle (\Delta {\bf{v}})^2 \rangle$ in this limit, let $f({\bf v })$ be any normalized distribution of velocities. Then,
\begin{align}
\langle (\Delta {\bf v} )^2\rangle &= \int({\bf v_2}-{\bf v_1})^2 f({\bf v_1}) f({\bf v_2}) d {\bf v_1} d {\bf v_2} \notag \\
&= 2\left(\langle v^2 \rangle-\langle {\bf v} \rangle ^2 \right) = 2 \sigma^2_{{\bf v}}.
\end{align}
Therefore,
\begin{align}
f_{\rm heur}(k) = \frac{\langle (\Delta {\bf{v}})^2 \rangle}{2 \sigma^2_{{\bf v}}}.
\label{fheur}
\end{align} 

We now calculate $\langle(\Delta {\bf v})^2\rangle$ for the position-averaged $\Delta {\bf v}$ for a single, divergenceless (i.e. transverse) mode with wavelength $\lambda$,
\begin{align}
{\bf v} = {\bf v_0} \sin\left(\frac{2 \pi}{\lambda} z\right).
\label{vel_field}
\end{align}
Let $P_{\Delta {\bf r}}(\Delta {\bf r})$ be the probability density that a
photon travels a displacement $\Delta {\bf r}$ between scatterings. Then, at a
given position ${\bf r}$,
\begin{align}
\langle (\Delta {\bf v})^2 \rangle_{{\bf r}} = \int \left( \Delta {\bf v} (\Delta {\bf r},{\bf r})\right)^2 P_{\Delta {\bf r}}(\Delta {\bf r}) d^3 \Delta {\bf r}. 
\end{align}
Averaging over all positions in a volume $V$, this is
\begin{align}
\langle (\Delta {\bf v})^2 \rangle = \frac{1}{V}\int \left( \Delta {\bf v}
(\Delta {\bf r},{\bf r})\right)^2 P_{\Delta {\bf r}}(\Delta {\bf r}) d^3
\Delta {\bf r} d^3 {\bf r}.
\end{align}
For a single mode, Eq. (\ref{vel_field}) gives
\begin{align}
\Delta {\bf v} = {\bf v_0} \sin\left(\frac{2 \pi}{\lambda}\left(z+
\Delta z\right)\right)-{\bf v_0}\sin\left(\frac{2 \pi}{\lambda} z\right)
\end{align}
and
\begin{align}
\langle (\Delta {\bf v})^2 \rangle = \frac{1}{\lambda}\int_0^\lambda dz
\int d^3\Delta{\bf r} \left(\Delta
{\bf v}(\Delta z,z)\right)^2 P_{\Delta{\bf r}}\left(\Delta{\bf r}\right).
\label{eq:dv2deltar}
\end{align}
The probability that a photon with mean free path $l$ travels a distance
between $s$ and $s+ds$ is $P_s\left(s\right)ds = (1/l)e^{-s/l} ds$.
Let $\mu = \cos \theta$, where $\theta$ is the angle between the photon
propagation direction and the $z$-axis, so that $\Delta z=s\mu$.
Then, expressing $\Delta{\bf r}$ in spherical polar coordinates and invoking
axisymmetry about the $z$-axis, equation (\ref{eq:dv2deltar}) becomes
\begin{align}
\langle (\Delta {\bf v})^2 \rangle = &\frac{v_0^2}{2l\lambda}\int_0^{\lambda}dz
\int_{-1}^1 {d\mu}\int_0^\infty ds
\left[\sin\left(\frac{2\pi}{\lambda}\left(z+s\mu\right)\right)\right.\notag \\
&-\left.\sin\left(\frac{2\pi}{\lambda}z\right)\right]^2 e^{-s/l}.
\end{align}
This is easily evaluated by performing the integral over $z$ first, giving
\begin{align}
\langle (\Delta {\bf v})^2 \rangle
%&= \frac{v_0^2}{l}\int_0^l \frac{1}{u}\int_0^\infty \left(1-
%\cos\left(\frac{2\pi}{\lambda}\Delta z\right)\right) e^{-\Delta z/u}
%d \Delta z du \\
%&= \frac{v_0^2}{l}\int_0^l \left(1-\frac{1}{1+\left(\frac{2\pi}{\lambda}u
%\right)^2}\right) du \\
%&= v_0^2\left(1 - \frac{1}{2\pi} \left(\frac{\lambda}{l}\right)\arctan
%\left(2 \pi \frac{l}{\lambda}\right)\right) \\
&= v_0^2\left[1 - \tau_k \tan^{-1}\left(\frac{1}{\tau_k}\right)\right]. \\
&= 2 \sigma_{{\bf v}}^2\left[1 - \tau_k \tan^{-1}\left(\frac{1}{\tau_k}\right)\right].
\end{align}
Eq. (\ref{fheur}) then gives
\begin{align}
f_{\rm heur}(k)=1 - \tau_k \tan^{-1}\left(\frac{1}{\tau_k}
\right).
\label{eq:heuristicsol}
\end{align}
The limiting cases are
\begin{align}
f_{\rm heur}(k)=\begin{cases}
1 &\mbox{if } \tau_k\rightarrow0\\
{1\over3\tau_k^2} &\mbox{if } \tau_k\rightarrow\infty.\end{cases}
\end{align}
By comparison, the exact solution for a single mode is determined by Eq. (\ref{eq:analyticsol}). Our heuristic result is plotted in Fig. \ref{MCdata1}, and is remarkably close to the exact solution. In particular, it is a much better approximation than the rough weighting function, Eq. (\ref{Twave3}), also shown in Fig. \ref{MCdata1}. Our model based on the second moment of the velocity difference distribution therefore captures the essential physics of Comptonization by a single mode.

Before proceeding to define $T_{\rm w, heur}(\lambda_{\rm p})$ for an arbitrary velocity field, we quantify the condition that the escape probability be low enough, presupposed by our derivation of $\langle(\Delta {\bf v})^2\rangle$. Since the distribution of velocity differences, as a function of
position, repeats every quarter wavelength, our results should be valid
provided photons travel a distance $\Delta z$ in the $z$ direction that is
greater than $\lambda_{\rm max}/4$ before escaping.  For an
(optically thick) random walk, $\Delta z \sim (N/3)^{1/2}\lambda_{\rm p}$,
where $N$ is the average number of scatterings. Since $N = 1/p_{\rm e} - 1$, where
$p_{\rm e}/t_{\rm C}$ is the escape probability per unit time during the average
time $t_{\rm C}$ between subsequent scattering events, 
\begin{align}
p_{\rm e} < \left(\frac{3}{16}\left(\lambda_{\rm max}/\lambda_{\rm p}\right)^2+1\right)^{-1}
\simeq\frac{16}{3}\left(\lambda_{\rm p}/\lambda_{\rm max}\right)^2
\end{align}
for optically thick modes. Up to a factor of unity, this agrees with Eq. (\ref{eq_escape_constraint1}).

We now use our model to compute $T_{\rm w, heur}(\lambda_{\rm p})$ for an arbitrary velocity field in terms of $T_{\rm w, heur}(\lambda_{\rm p}, k_i)$. Proceeding analogously to the single mode case, we define
\begin{align}
T_{\rm w, heur}(\lambda_{\rm p}) = \frac{\langle (\Delta {\bf{v}})^2 \rangle}{2 \sigma_{\bf v}^2} T_{\rm tot},
\label{Theur}
\end{align}
where $T_{\rm tot} = \sum_i T_{\rm tot,i}$ is the temperature corresponding to the full electron energy density of the velocity field. To simplify this, we compute
\begin{align}
\langle (\Delta {\bf v})^2 \rangle &= \frac{1}{V}\int\left(\sum_i\Delta{\bf v}_i
(\Delta {\bf r},{\bf r})\right)^2 P_{\Delta {\bf r}}(\Delta {\bf r}) d^3
\Delta {\bf r} d^3 {\bf r} \notag\\
&=\frac{1}{V}\int \left[\sum_i ({\bf v}_i({\bf r}+\Delta {\bf r})-{\bf v}_i
({\bf r}))\right]^2 P_{\Delta {\bf r}}(\Delta {\bf r}) d^3 \Delta {\bf r}
d^3 {\bf r} \notag\\
&=\frac{1}{V}\int \sum_i \left({\bf v}_i({\bf r}+\Delta {\bf r})-{\bf v}_i
({\bf r})\right)^2 P_{\Delta {\bf r}}(\Delta {\bf r}) d^3 \Delta {\bf r} d^3
{\bf r} \notag\\
%&=\frac{1}{V}\int \sum_i \left( \Delta {\bf v}_i(\Delta {\bf r},{\bf r})
%\right)^2 P_{\Delta {\bf r}}(\Delta {\bf r}) d^3 \Delta {\bf r}
%d^3 {\bf r}\notag \\
&= \sum_i \langle (\Delta {\bf v}_i)^2 \rangle. 
\label{deltavcalc}
\end{align}
To get from line 2 to line 3 we made use of the orthogonality for distinct sinusoidal modes. That is, for two distinct modes, ${\bf v_i}$ and ${\bf v_j}$, $i \neq j$, and any displacement ${\Delta \bf r}$,
\begin{align}
\int {\bf v}_i ({\bf r} + \Delta {\bf r}) \cdot {\bf v}_j ({\bf r})d^3
{\bf r} = 0.
\end{align}
Then, since $T_{\rm tot,i} \propto \sigma_{{\bf v_i}}^2$, Eqs. (\ref{fheur}), (\ref{Theur}), and (\ref{deltavcalc}) give
\begin{align}
T_{\rm w, heur}(\lambda_{\rm p}) = \sum_i f_{\rm heur}(k_i)T_{\rm tot,i},
\label{Theur2}
\end{align}
or, alternatively,
\begin{align}
T_{\rm w, heur}(\lambda_{\rm p}) = \sum_i T_{\rm w, heur}(\lambda_{\rm p},k_i).
\end{align}
Eq. (\ref{Theur2}) is the same as the exact solution for the wave temperature, Eq. (\ref{Twave2}), except that here the heuristic weighting function is used. Note that in our heuristic derivation it is the orthogonality of distinct modes that allows us to express the wave temperature of an arbitrary velocity field as a sum over the wave temperatures of its modes. Unsurprisingly, orthogonality was used analogously to prove Eq. (\ref{Twave2}). Therefore, our model based on the second moment of the velocity difference distribution captures the essential physics of Comptonization by an arbitrary, divergenceless velocity field.

\subsection{The dependence of the wave temperature on the turbulence power spectrum}
\label{sec_spec}
We now analyse the dependence of $T_{\rm w}(\lambda_{\rm p})$ on the power spectrum of the turbulence. For the remainder of this section, we write $k$ in units of $1/\lambda_{\rm p}$ for clarity (i.e. $k \ll 1$ and $k \gg 1$ denote optically thick and thin scales, respectively). If the turbulence is completely optically thin on all scales ($k_{\rm min} \gg 1$), then $T_{\rm w}(\lambda_{\rm p})=T_{\rm tot}$, independent of the energy spectral index, $p$.  However, if some scales in the
turbulent cascade are optically thick, then $p$ will affect $T_{\rm w}(\lambda_{\rm p})$. 

\begin{figure}
\includegraphics[width = 84mm]{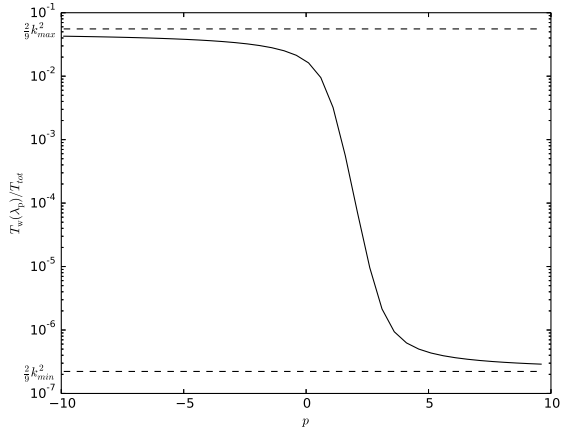}
\caption{The dependence of $T_{\rm w}(\lambda_{\rm p})/T_{\rm tot}$
on $p$ (calculated exactly) for $k_{\rm min} = 0.001$, $k_{\rm max} = 0.5$, where
$k$ is in units of $1/\lambda_{\rm p}$. Note that $T_{\rm w}(\lambda_{\rm p})/T_{\rm tot}$
approaches $\frac{2}{9} k_{\rm max}^2$ and $\frac{2}{9} k_{\rm min}^2$
for $p \ll 1$ and $p \gg 3$, respectively, as expected for a broad power
spectrum with $k_{\rm max} < 1$.}
\label{Ueff_p}
\end{figure}

For the case where all scales in the turbulence are optically thick ($k_{\rm max} \ll 1$), Eq. (\ref{eq:analyticsol}) implies that
\begin{align}
f(k) = \frac{2}{9} k^2.
\label{optically_thick_weight}
\end{align}
Integrating this over an energy spectrum $T_{\rm tot}(k) \propto k^{-p}$ then gives
\begin{align}
\frac{T_{\rm w}(\lambda_{\rm p})}{T_{\rm tot}} =\left(\frac{2}{9}\right)\frac{1-p}{3-p}
\left(\frac{k_{\rm max}^{3-p} - k_{\rm min}^{3-p}}{k_{\rm max}^{1-p} -
k_{\rm min}^{1-p}}\right).
\end{align}
For a broad power spectrum, i.e. $k_{\rm min}/k_{\rm max} \ll 1$,
this simplifies to
\begin{align}
\frac{T_{\rm w}(\lambda_{\rm p})}{T_{\rm tot}} = \frac{1}{9}\times
\begin{cases}
2k_{\rm max}^2, & p \ll 1\\
k_{\rm max}^{4/3}k_{\rm min}^{2/3}, & p=5/3\\
2k_{\rm min}^2, & p \gg 3.
\end{cases}
\end{align}
This is illustrated in Fig. \ref{Ueff_p}. Note that
$T_{\rm w}(\lambda_{\rm p})/T_{\rm tot}$ drops significantly for $p > 1$, because then
the energy bearing modes are on the largest scales.  These are the most
optically thick and therefore the most downweighted in their contribution
to bulk Comptonization.

We next analyse whether $T_{\rm w}(\lambda_{\rm p})/T_{\rm tot}$, for a given spectral
index $p$ and range of modes, $k_{\rm min} < k < k_{\rm max}$, is dominated
by small or large scales. In other words, we examine which turbulent modes
in a given spectrum contribute most to bulk Comptonization. The relative
contribution of a scale with wavenumber $k$ is
\begin{align}
T_{\rm w}(\lambda_{\rm p},k)dk &\sim T_{\rm w}(\lambda_{\rm p},k) k \notag\\
&\sim f(k) T_{\rm tot}(k) k \notag\\
&\sim k^{q-p+1},
\label{contribution}
\end{align}
where, from Eq. (\ref{eq:analyticsol}), $q=2$ for optically thick
($k\ll1$) scales, and $q=0$ for optically thin ($k\gg1$) scales. Now consider
an underlying power spectrum with some $k_{\rm min}$ and $k_{\rm max}$.
We see that for $p < 1$ the exponent in Eq. (\ref{contribution}) is
always positive, and so small scales contribute most to bulk Comptonization,
regardless of $k_{\rm min}$ and $k_{\rm max}$. This is physically intuitive;
for $p < 1$, the turbulent power is concentrated on small scales. Since the
weighting factor $f(k)$ also favors small scales, they of course contribute
most. For $p > 3$, the exponent is always negative, and so large scales always
contribute most. In this case, the turbulent power is so concentrated on large
scales that they contribute more even though $f(k)$ favors small scales. 

For $1 < p < 3$, we first consider the part of the spectrum with $k \gg 1$
(if it exists). Since $q=0$, small scales contribute more than large scales
for these modes. Now consider the part of the spectrum with $k \ll 1$ (if it
exists). Here, large scales contribute more than small scales. Therefore,
looking at the entire power spectrum, it is intermediate scales that contribute
most, assuming it is broad enough to include regions of both small and large
$k$. If it is not sufficiently broad, then whether small or large scales
contribute most depends on $k_{\rm min}$ and $k_{\rm max}$ relative to
$k \approx 1$ (the optically thin to thick transition wavenumber).

\begin{figure}
\includegraphics[width = 84mm]{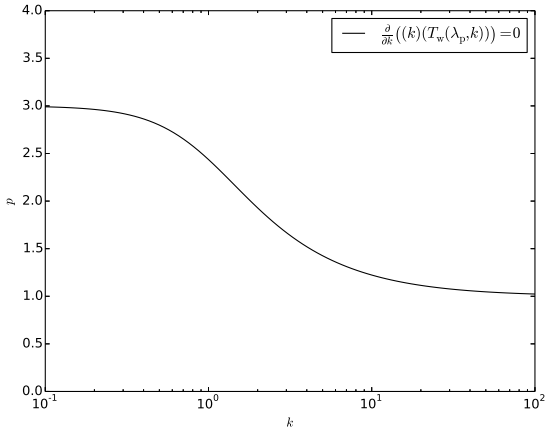}
\caption{Regions in $k-p$ space in which bulk Comptonization is dominated by
small (lower region) and large (upper region) scales, as determined by the
sign of $\frac{d}{d k}((k)(T_{\rm w}(\lambda_{\rm p},k)))$.}
\label{Ueff_dist}
\end{figure}

These results are depicted in Fig. \ref{Ueff_dist}.  The curve in this figure
shows the values of $p$ and $k$ such that the derivative of
$kT_{\rm w}(\lambda_{\rm p},k)=kf(k)T_{\rm tot}(k)$ is zero, using the full analytic
expression for $f(k)$ from Eq. (\ref{eq:analyticsol}).  To connect this
figure to our discussion, draw a horizontal line from $k_{\rm min}$ to
$k_{\rm max}$ at a given value of $p$. If the line lies in the lower (upper)
region, then small (large) scales contribute most. If the line straddles the
two regions, then for the part of the line that lies in the lower region small
scales contribute most, and for the part that lies in the upper region large
scales contribute the most. In this case, then, for the entire spectrum it is
the scales that straddle the curve which contribute most. Note that for
$p < 1$ and $p > 3$ a spectrum can never straddle the curve, whereas for a
Kolmogorov spectrum ($p = 5/3$), e.g., it can, if $k_{\rm min} < 3 < k_{\rm max}$.

\section{Comptonization by turbulence with non-zero divergence}
\label{sec_compressible}
\cite{soc04} conjectured that Comptonization by turbulence can be treated as thermal Comptonization by solving the Kompaneets equation with an equivalent ``wave" temperature. In section \ref{sec_incompressible} we proved this under certain conditions, one of which is that the turbulence be divergenceless. In this section we investigate Comptonization by velocity fields with non-zero divergence, an effect that usually cannot be treated as thermal Comptonization, with the aim of understanding how it impacts radiation spectra in generic, turbulent regions of stratified accretion disc atmospheres.

While we do not have an exact solution for Comptonization by an arbitrary velocity field with non-zero divergence, we have solutions for two limiting cases. The trivial case is the optically thin case, i.e. when the mean free path is significantly larger than the largest length scale in the turbulence, $\lambda_{\rm p} / \lambda_{\rm max} \gg 1$. Electron velocities ``sampled" by photons are uncorrelated and so Compton scattering should depend on only the total spatial average distribution of electron velocities. This is, therefore, the one case where Comptonization by a velocity field with non-zero divergence can be treated as thermal Comptonization, by solving the Kompaneets equation with $T_{\rm w}(\lambda_{\rm p})$ given by
\begin{align}
\frac{3}{2} k_{\rm B} T_{\rm w}(\lambda_{\rm p}) = \frac{3}{2} k_{\rm B} T_{\rm tot} = \frac{1}{2} m \left\langle v^2 \right\rangle.
\label{compressible_thin}
\end{align}
Note that although we formally write $T_{\rm w}(\lambda_{\rm p})$ as a function of $\lambda_{\rm p}$, in this limit $T_{\rm w}(\lambda_{\rm p})$ is independent of $\lambda_{\rm p}$. Energy exchange first order in velocity vanishes since photons are equally likely to downscatter as they are to upscatter. Bulk Comptonization is then solely due to radiation viscous dissipation (see section \ref{sec_general}). As viscous effects are second order in velocity, it is unsurprising that they can be characterized by a temperature. We also note that in this limit the wave temperature is the same as that for a divergenceless velocity field, Eq. (\ref{eq_limits}). Optically thin bulk Comptonization is therefore a single phenomenon that depends on only the mean square speed of the velocity field.

To arrive at this result with a more formal approach, we start with the zeroth moment of the radiative transfer equation, Eq. (\ref{eqpsaltis}).
In the limit of optically thin turbulence the radiation variables must be homogeneous and isotropic, so that $n^i=0=n^{ijk}$, and $n^{ij}=(1/3)n\delta^{ij}$. Then, averaging Eq. (\ref{eqpsaltis}) over the largest scale $\lambda_{\rm max}$ gives
\begin{eqnarray}
\frac{\lambda_{\rm p}}{c}{\partial n\over\partial t}=
{1\over mc^2\epsilon^2}{\partial\over\partial\epsilon}
\Biggl\{\epsilon^4\Biggl[n+n^2+\cr
\left(k_{\rm B}T_{\rm e}+{1\over3}m\langle v^2\rangle\right)
{\partial\over\partial\epsilon}n\Biggr]\Biggr\}.
\label{eqkomp}
\end{eqnarray}
This is the Kompaneets equation, with the contribution from the velocity field to the Comptonization temperature given by Eq. (\ref{compressible_thin}). 

In the optically thick case, i.e. when the photon mean free path is significantly smaller than the smallest scale in the turbulence, the lowest order energy exchange is the work done by radiation pressure to compress the gas, since it is first order in velocity and since radiation viscous effects are suppressed by a factor $\lambda_{\rm p}^2/\lambda^2$ (section \ref{sec_incompressible}). We focus on the extremely optically thick case, which we define as the limit in which photon diffusion is negligible relative to photon advection, so that the photon and gas fluids are strongly coupled. If we define $\psi_\epsilon = \epsilon^2 n$, the photon number density at energy $\epsilon$, and $\psi = \int \psi_\epsilon d \epsilon$, the total photon number density, then the advection and diffusion fluxes are given by
\begin{align}
{\bf F_{\rm a}} = {\bf v}\psi
\end{align}
and
\begin{align}
{\bf F_{\rm d}} = - \frac{1}{3}\lambda_{\rm p}c {\bf \nabla} \psi,
\end{align}
respectively. The extremely optically thick limit is then given by
\begin{align}
\frac{\lambda_{\rm min}}{\lambda_{\rm p}} \gg \frac{c}{v}.
\label{eq_extremely_optically_thick}
\end{align}
In this case, velocity convergence corresponds to compression in which gas mechanical energy is transferred locally to the photons. We expect that photons with wavelength $\lambda_{\rm \gamma}$ are effectively compressed at a rate given by the velocity difference across $\lambda_{\rm \gamma}$, 
\begin{align}
\frac{d \lambda_{\rm \gamma}}{d t} = \frac{1}{3}\lambda_{\rm \gamma} {\bf \nabla} \cdot {\bf v},
\label{eq_photon_compression}
\end{align}
so that, for example, a locally thermal photon distribution remains thermal and only changes temperature, completely analogous to the evolution of the cosmic microwave background radiation under the expansion of the universe.  This is equivalent to a fractional energy change per scattering given by
\begin{align}
\frac{\lambda_{\rm p}}{c} \frac{1}{\epsilon}\frac{d \epsilon}{d t} = - \frac{\lambda_{\rm p}{\bf \nabla} \cdot {\bf v}}{3c} ,
\label{upscattering_fraction}
\end{align}
since 
\begin{align}
\frac{d\epsilon}{d \lambda_{\rm \gamma}} = - \frac{\epsilon}{\lambda_{\rm \gamma}}.
\end{align}
We now confirm that Eq. (\ref{upscattering_fraction}) correctly describes extremely optically thick Compton scattering in a converging or diverging flow, both by providing a heuristic argument and by deriving it from the radiative transfer equation.

Before proceeding, we note that the study of photon upscattering by a converging velocity field can be traced back to \cite{bla81a,bla81b} and \cite{pay81}, who, along with later authors, made detailed spectral calculations for specific velocity fields in shocks and spherically accreting systems. In fact, Eq. (\ref{upscattering_fraction}) can equivalently be stated as the upscattering timescale 
\begin{align}
t_{\rm up}^{-1} = \frac{1}{3}{\bf \nabla} \cdot {\bf v}
\end{align}
given in \cite{bla81a}. In this section, by contrast, we have been investigating how this effect manifests itself locally in a generic, turbulent region of a stratified disc atmosphere, with the goal of resolving and interpreting it in spectral calculations of radiation MHD simulations. We have been focusing on the extremely optically thick case, in which the photon and gas fluids are strongly coupled, because the physics is both relevant and intuitive. In the moderately optically thick case, on the other hand, i.e.
\begin{align}
1 \ll \frac{\lambda_{\rm min}}{\lambda_{\rm p}} \sim \frac{c}{v},
\end{align}
such as photon upscattering in a radiation pressure dominated shock \citep{bla81b}, diffusion competes with advection so that photon distributions at neighboring fluid elements mix. Photon upscattering in such a converging flow may not be viewed as simply the compression of a photon fluid strongly coupled to the gas, and photon upscattering in which a photon thermal distribution is not preserved can occur.

To heuristically derive Eq. (\ref{upscattering_fraction}), consider a disturbance converging in the $\hat{z}$ direction given by 
\begin{align}
{\bf v}  = -\alpha z \hat{z}.
\label{v_func}
\end{align}
For a 3D random walk, the average distance between scatterings traveled by a photon in the direction of convergence is $\lambda_{\rm p}/3$. Since the fractional energy change per scattering for low energy photons is $v/c$, at $z = 0$ this gives
\begin{align}
\frac{\Delta \epsilon}{\epsilon} = \frac{-\lambda_{\rm p} \partial_z v_z}{3c},
\end{align}
in agreement with Eq. (\ref{upscattering_fraction}).

We now derive Eq. (\ref{upscattering_fraction}) with the radiative transfer equation, Eq. (\ref{eqpsaltis}). If we (1) omit stimulated scattering terms to facilitate comparison with simulations, (2) substitute in the standard closure relations for the first moment in the optically thick limit,
\begin{align}
n^i = -\frac{v_i}{2c} \epsilon \frac{\partial n}{\partial \epsilon} -
\frac{1}{3}\lambda_{\rm p} \partial_i n 
\label{closure1}
\end{align}
and
\begin{align}
n^{ij} = \frac{1}{3}n \delta^{ij},
\end{align}
and (3) substitute in the photon number density $\psi_{\epsilon}$, then the radiative transfer equation to second order in velocity and first order in $\epsilon / mc^2$ becomes
\begin{align}
\frac{\lambda_{\rm p}}{c} \partial_{t} \psi_{\epsilon} &= -\frac{\lambda_{\rm p}}{c} {\bf \nabla}\cdot \left({\bf v} \psi_{\epsilon} - \frac{1}{3}\lambda_{\rm p}c {\bf \nabla} \psi_{\epsilon}\right) \notag \\
& - \partial_\epsilon\left( \epsilon \left(\frac{-\lambda_{\rm p} {\bf \nabla} \cdot {\bf v}}{3c} \right)
\psi_{\epsilon}\right)
- \partial_\epsilon\left( \epsilon
\left(\frac{4 k_{\rm B} T_{\rm e} - \epsilon}{mc^2}\right)\psi_{\epsilon}\right) \notag \\
&+ \partial_\epsilon^2 \left(\epsilon^2 \left(\frac{k_{\rm B} T_{\rm e}}{m c^2}\right)\psi_\epsilon \right).
\label{2D_diff}
\end{align}
Neglecting stimulated scattering, this is \cite{bla81a} Eq. (18), cast in the physically revealing form of a Fokker-Plank equation. The terms inside the divergence operator correspond to spatial drift (i.e. photon advection) and spatial diffusion, respectively. The next term corresponds to energy drift due to photon upscattering (downscattering) in the presence of a converging (diverging) velocity field. The remaining terms correspond to energy drift and diffusion due to thermal Comptonization. Note that even though  \cite{bla81a} start with a zeroth moment equation correct only to first order in $v/c$, their resulting equation is the same because with the standard closure relation (Eq. \ref{closure1}) the second order terms cancel. The fractional energy change per scattering given by Eq. (\ref{upscattering_fraction}) follows from the bulk upscattering term. In the extremely optically thick limit the spatial diffusion term is negligible, and so photons are advected with the velocity field and upscatter according to Eq. (\ref{upscattering_fraction}). The effect of this process on the emergent spectrum of the disc will depend primarily on how effectively photons are able to escape from converging (or diverging) regions to the observer.

\section{Discussion}
\label{sec_discussion}
We have investigated how photon spectra produced by turbulent Comptonization depend on the properties of the turbulence itself, and how to resolve and interpret this effect in radiation MHD simulations. Our principle results are as follows. First we address the conjecture of \cite{soc04} that turbulent Comptonization can be treated as thermal Componization by solving the Kompaneets equation with an equivalent ``wave" temperature. For Comptonization by divergenceless turbulence, which is due to the radiation viscous shear stress tensor and is therefore an effect second order in velocity, this conjecture holds provided that the density is uniform and the escape probability is sufficiently small (Eq. \ref{eq_escape_constraint1}). The correct wave temperature can be expressed in terms of the radiation viscous shear stress tensor (Eqs. \ref{eq_Twave_n} and \ref{eq_Twave_P}) and is in turn given by Eq. (\ref{Twave2}). This is a weighting over the velocity power at each scale in the turbulence, with the weighting function given by Eq. (\ref{eq:analyticsol}). Optically thin scales have unit weight factors, implying that the power at these scales contributes fully to the wave temperature. Optically thick scales, on the other hand, are downweighted by a factor proportional to one over the square of the optical depth across those scales. This downweighting arises because bulk Comptonization is sensitive to the difference in bulk fluid velocity between subsequent scatterings, as demonstrated by our heuristic argument in section \ref{sec_incompressible}. These velocity differences are much smaller than the velocity amplitude on some scale if the photon mean free path is much smaller than that scale.

We note that our expression for the wave temperature of divergenceless turbulence assumes a uniform density. Furthermore, it is non-local in that it is expressed in terms of the mode decomposition of the entire velocity field. Although a generic velocity field has density variations, if the largest turbulent length scale $\lambda_{\rm max}$ is smaller than the scale on which the density varies then we can define the wave temperature for regions of order $\lambda_{\rm max}$. If the density also varies on smaller scales, then we can define the wave temperature for smaller regions by omitting contributions from larger modes. Assuming these modes would be significantly downweighted anyway, this procedure would likely capture nearly all Comptonized power in the turbulence.

For turbulence with non-zero divergence, bulk Comptonization energy exchange is due to both work done by radiation pressure and radiation viscous dissipation. Since pressure work is an effect first order in velocity, it cannot be treated with an equivalent temperature. But in the optically thin limit, i.e. when the mean free path is significantly larger than the largest length scale in the turbulence, energy exchange first order in velocity vanishes since photons are equally likely to downscatter as they are to upscatter. Bulk Comptonization by a velocity field with non-zero divergence is then solely due to radiation viscous dissipation, and it may be treated as thermal Comptonization. The wave temperature is given by the full power in the turbulence, Eq. (\ref{compressible_thin}), directly analogous to the optically thin limit of divergenceless turbulence.

In the optically thick case, i.e. when the mean free path is significantly smaller than the smallest scale in the turbulence, the lowest order energy exchange is the work done by radiation pressure to compress the gas, since it is first order in velocity and since radiation viscous effects are suppressed by a factor $\lambda_{\rm p}^2/\lambda^2$. The extremely optically thick limit, defined as the limit in which spatial photon diffusion is negligible relative to photon advection, is given by Eq. (\ref{eq_extremely_optically_thick}). In this limit, the photon and gas fluids are strongly coupled, so that velocity convergence corresponds to compression in which gas mechanical energy is transferred locally to the photons. Photons upscatter with fractional energy change given by Eq. (\ref{upscattering_fraction}). In this process a locally thermal photon distribution remains thermal and only changes temperature, analogous to the evolution of the cosmic microwave background radiation under the expansion of the universe. Unlike energy exchange due to viscous dissipation, this process is reversible. The effect of this process on the emergent spectrum of the disc will depend primarily on how effectively photons are able to escape from such regions to the observer.

For moderate optical depths, such as photon upscattering in a radiation pressure dominated shock \citep{bla81b}, diffusion competes with advection so that photon distributions at neighboring fluid elements mix. The effect of photon upscattering in a converging region on local spectra is more complex and depends on the details of the flow.

It is important to note that optically thick Comptonization by a velocity field with non-zero divergence may be very sensitive to the time dependence of the velocity field. This is important because usually post-processing Monte-Carlo simulations invoke time-independent atmospheres with the assumption that they approximately capture the effects of interest. For example, consider Comptonization in a converging region. A time-independent velocity field results in the accumulation of photons and subsequent upscattering to high energies at the point of zero velocity. If the region is near the photosphere, the emergent spectrum will be strongly upscattered. But in a time-dependent velocity field photons have a limited time to upscatter before converging regions become diverging regions, and so the effect on the emergent spectrum is significantly different.

Real accretion flows are spatially stratified. Simulations of magnetorotational turbulence generally indicate that this turbulence dominates the fluid velocities in the midplane regions of the accretion flow.  This turbulence is largely incompressible (divergenceless), although it generally excites compressible spiral acoustic waves \citep{hei09a,hei09b}.  Sufficiently far from the midplane, magnetic forces always dominate thermal pressure forces, and support the flow vertically against the tidal gravitational field of the compact object.  Such regions are dominated by Parker instability dynamics, and exhibit considerable compressive behavior (i.e. the flow has non-zero divergence) with large density fluctuations \citep{bla07}.

In order for radiation MHD simulations to properly account for energy exchange due to turbulent Comptonization so that post-processing Monte Carlo simulations of photon spectra are self-consistent, they must include energy terms second order in velocity and use a moment closure scheme that correctly captures contributions to the radiation stress tensor that are first order in velocity. The appropriate energy equation source terms in lab and fluid frame variables are given by Eqs. (\ref{integrated1}) and (\ref{eq_intergrated2}), respectively. Flux-limited diffusion and the M1 closure scheme are insufficient because they neglect the lowest order contribution to the radiation stress tensor. In this case, instead of exchanging energy with photons, turbulence will eventually cascade down to the gridscale (or viscous or resistive scale if the code has explicit viscosity or resistivity) and increase the internal energy of the gas. Gas internal energy may then be exchanged with photons through thermal Comptonization. The omission of viscous dissipation by radiation therefore does not prevent the eventual transfer of turbulent energy to radiation, but it may have other physical effects which can in turn affect radiation spectra.

\section{Conclusions}
\label{sec_conclusion}
Radiation pressure dominated accretion flows may have bulk velocities that exceed thermal electron velocities. Turbulent Comptonization, which is the effect of Comptonization by bulk, turbulent electron motions on photon spectra, should be significant in such sources.

For turbulence with zero divergence, bulk Comptonization is equivalent to a radiation viscous shear stress tensor acting on the fluid. The resulting Comptonized spectrum can be calculated by solving the Kompaneets equation with an equivalent ``wave" temperature, which is a weighted sum over the power present at each scale in the turbulent cascade (Eq. \ref{Twave2}).

Bulk Comptonization by turbulence with non-zero divergence is due to both pressure work and radiation viscous dissipation. In the limit of optically thin turbulence, pressure work has negligible effect on photon spectra and radiation viscous dissipation alone can be treated as thermal Comptonization with a temperature equivalent to the full turbulent power (Eq. \ref{compressible_thin}). In the limit of extremely optically thick turbulence (Eq. \ref{eq_extremely_optically_thick}), radiation viscous dissipation is suppressed, and the evolution of local photon spectra can be understood in terms of compression and expansion of the strongly coupled photon and gas fluids. Photons upscatter with fractional energy change given by Eq. (\ref{upscattering_fraction}). In this process a locally thermal photon distribution remains thermal and only changes temperature. The effect of this process on the emergent spectrum of the disc will depend primarily on how effectively photons are able to escape from such regions to the observer.

Modeling turbulent Comptonization will ultimately require detailed analysis of radiation MHD simulations and post-processing Monte-Carlo simulations. By exploring how photon spectra produced by turbulent Comptonization depend on the properties of the turbulence itself, we have laid the groundwork necessary to make sure this effect is both captured and correctly interpreted in these simulations.

\section*{Acknowledgements}
We thank Aristotle Socrates for useful
discussions.
We also thank Donald Pakey for forwarding errata to the paper by \citet{psa97},
and Li Xue and Feng Yuan for hosting OB at Xiamen University and Shanghai
Astronomical Observatory where part of this work was carried out.
We gratefully acknowledge support by NASA Astrophysics Theory
Program grant NNX13AG61G, and the International Space Science Institute (ISSI)
in Bern.

\onecolumn
\appendix
\section{Derivation of the occupation number second moment due to an arbitrary, divergenceless velocity field of uniform density to first order in velocity}
\label{transverse_solution}

Beginning with Eq. (\ref{eq_transfer_first_order}), we prove that the steady state occupation number second moment for a divergenceless velocity field of uniform density to first order in velocity is given by Eq. (\ref{eq_nij}). First we find the solution for a single mode given by
\begin{align}
{\bf v} = \sqrt{2}v_{\rm rms} \sin\left(\frac{2\pi z}{\lambda}\right){\bf \hat{x}} = v(z) {\bf \hat{x}}.
\label{incompressible_v2}
\end{align}
For this mode, the transfer equation is
\begin{align}
\lambda_{\rm p} \ell^z \partial_z n_1({\bf \hat{\ell}},z) = -n_1({\bf \hat{\ell}},z) - \ell^x v(z) \epsilon\partial_{\epsilon}n_0 + \frac{3}{2}\ell^x \ell^z n_{1}^{xz},
\end{align}
where we assume $n_1=0$ and $n_1^{zz} = 0$, which we
can check later. Then, the transfer equation is
\begin{equation}
\lambda_{\rm p} \ell^z{\partial n_1
\over\partial z}=-n_1
-\ell^x{\sqrt{2}v_{\rm rms}}\sin\left({2\pi z\over\lambda}\right)
\epsilon{\partial n_0\over\partial\epsilon}
+{3\over2}\ell^x\ell^z{1\over4\pi}\oint d\Omega^\prime \ell^{\prime x} \ell^{\prime z}
n_1(\hat{\bf \ell}^\prime,z).
\label{eq:integrodiffinc}
\end{equation}
First we address the $z$-dependence. Because this equation is linear, it must be
that $n_1({\bf \hat{\ell}},z)$ is a superposition of a sine and a cosine,
\begin{equation}
n_1(\hat{\bf \ell},z)
=A(\hat{\bf \ell})\cos\left({2\pi z\over\lambda}\right)
+B(\hat{\bf \ell})\sin\left({2\pi z\over\lambda}\right).
\end{equation}
This gives two coupled integral equations for $A$ and $B$,
\begin{equation}
-{\ell^z\over\tau_k}A=-B-\ell^x{\sqrt{2}v_{\rm rms}}\epsilon{\partial n_0\over\partial\epsilon}
+{3\over2} \ell^x \ell^z{1\over4\pi}\oint d\Omega^\prime \ell^{\prime x} \ell^{\prime z}
B(\hat{\bf \ell}^\prime)
\end{equation}
and
\begin{equation}
{\ell^z\over\tau_k}B=-A
+{3\over2} \ell^x\ell^z{1\over4\pi}\oint d\Omega^\prime \ell^{\prime x} \ell^{\prime z}
A(\hat{\bf \ell}^\prime).
\end{equation}
It seems that both $A$ and $B$ are proportional
to one power of $\ell^x$.
Writing $A=\ell^x\tilde{A}$ and $B=\ell^x\tilde{B}$, and $\ell^z=\cos\theta=\mu$,
we then obtain
\begin{equation}
-{\mu\over\tau_k}\tilde{A}=-\tilde{B}-
{\sqrt{2}v_{\rm rms}}\epsilon{\partial n_0\over\partial \epsilon}
+{3\over8}\mu\int_{-1}^1d\mu^\prime(1-{\mu^\prime}^2)\mu^\prime\tilde{B}
(\mu^\prime)
\end{equation}
and
\begin{equation}
{\mu\over\tau_k}\tilde{B}=-\tilde{A}
+{3\over8}\mu\int_{-1}^1d\mu^\prime(1-{\mu^\prime}^2)\mu^\prime\tilde{A}
(\mu^\prime).
\end{equation}
Observing that letting $\tilde{A}$ and $\tilde{B}$ be odd and even, respectively, is a consistent solution, the $\mu^\prime$ integral of
$\tilde{B}$ vanishes, and the two equations can be combined to give a single
equation for $\tilde{A}$,
\begin{equation}
-{\mu\over\tau_k}{\sqrt{2}v_{\rm rms}}\epsilon{\partial n_0 \over\partial \epsilon}
+{\mu^2\over\tau_k^2}\tilde{A}=-\tilde{A}
+{3\over8}\mu\int_{-1}^1d\mu^\prime(1-{\mu^\prime}^2)\mu^\prime\tilde{A}
(\mu^\prime).
\end{equation}
We then solve this equation with a series expansion of odd powers
of $\mu$:
\begin{equation}
\tilde{A}=\sum_{n=0}^\infty a_{2n+1}\mu^{2n+1}
\end{equation}
gives
\begin{equation}
a_{2n+1}=(-1)^n{a_1\over\tau_k^{2n}},
\end{equation}
which is just the expansion of $(1+\mu^2/\tau_k^2)^{-1}$, so that
\begin{equation}
\tilde{A}={a_1\mu\over1+\mu^2/\tau_k^2}.
\end{equation}
Substituting this back into the integral equation gives $a_1$, which completes
the solution. So far then, we have
\begin{eqnarray}
n_1(\hat{\bf \ell},z)&={\sqrt{2}v_{\rm rms}}\epsilon{\partial n_0\over
\partial\epsilon}\sin\theta\cos\phi\Biggl\{\left({1\over Q}\right)
{\tau_k^2\cos\theta\over\tau_k^2+\cos^2\theta}\cos\left({2\pi z\over\lambda}
\right) +\left[{\tau_k\cos^2\theta\over Q(\tau_k^2+\cos^2\theta)}-1\right]\sin
\left({2\pi z\over\lambda}\right)\Biggr\},
\end{eqnarray}
where
\begin{equation}
Q\equiv\tau_k-{3\over4}\tau_k^3\int_0^1d\mu{\mu^2-\mu^4\over\tau_k^2+\mu^2}
=\tau_k-{3\over4}\tau_k^3\left[{2\over3}+\tau_k^2-\tau_k(1+\tau_k^2)
\tan^{-1}\left({1\over\tau_k}\right)\right].
\label{eq:qexact}
\end{equation}
Note that $Q\rightarrow\tau_k-\tau_k^3/2$ in the optically thin limit,
and $Q\rightarrow9\tau_k/10$ in the optically thick limit. This solution is consistent with our assumptions and solves Eq. (\ref{eq_transfer_first_order}) to first order in velocity. The second moment is
\begin{align}
n^{ij} = \frac{1}{3}n_0 \delta^{ij}+ \frac{\lambda_{\rm p} \epsilon \partial_\epsilon n_0}{3c} \tau_k^2  f(\tau_k) \left( \partial_i v^j + \partial_j v^i\right),
\label{eq_second_moment1}
\end{align}
where $f(\tau_k)$ is given by Eq. (\ref{eq:analyticsol}). Since Eq. (\ref{eq_transfer_first_order}) is linear, the solution for an arbitrary, divergenceless velocity field of uniform density is then given by Eq. (\ref{eq_nij}).

\label{lastpage}

\end{document}